\documentclass[reprint, superscriptaddress, longbibliography, nofootinbib,
 amsmath,amssymb,
 aps,
]{revtex4-2}

\usepackage{graphicx}

\usepackage{bm}

\usepackage{hyperref}

\usepackage{physics}
\usepackage{orcidlink}

\begin{document}

\title{How many photons can a molecule scatter before its coherence is lost?}

\author{M.~R.~Tarbutt\orcidlink{0000-0003-2713-9531}}
\email{m.tarbutt@imperial.ac.uk}
\affiliation{Centre for Cold Matter, Blackett Laboratory, Imperial College London, London SW7 2AZ, United Kingdom}

\begin{abstract}
Molecular qubits held in conservative optical traps may lose coherence when they scatter photons from the trapping lasers. The rate for this process has two parts, one equal to the mean Raman scattering rate of the two qubit states, and the other proportional to the square of the difference in their Rayleigh scattering amplitudes. This paper calculates the decoherence rate, expressing results in terms of the vector and tensor parts of the molecular polarizability, $\alpha_1$ and $\alpha_2$. While the decoherence rate is generally much larger for molecules than for atoms, it is suppressed at wavelengths where $\alpha_1$ and $\alpha_2$ are both small. The wavelengths where this occurs are identified for CaF molecules.
\end{abstract}

\maketitle

\section{Introduction}

Ultracold molecules are increasingly being used to test fundamental physics~\cite{DeMille2024, Anderegg2023, Zeng2024, Lasner2025, Jenkins2026, Barontini2022Short}, study strongly-interacting quantum gases~\cite{Li2023, Duda2023, Schindewolf2026, Zhang2026}, understand reactions in the fully quantum regime~\cite{Liu2024,Liu2025}, process quantum information~\cite{Bao2023, Holland2023, Picard2024, Holland2025, Yu2026_dipole, Ruttley2025}, and simulate a wide variety of many-body quantum systems~\cite{Cornish2024, Holland2026arxiv, Raghuram2026arxiv}. These applications rely on long-lived quantum coherence of the molecules. The experiments are usually done with molecules in optical traps, so photon scattering may contribute to decoherence. We are especially interested in coherences within the molecule, rather than their motional states. Raman scattering, where the final internal state differs from the initial one, produces entanglement between the molecule and the photon, resulting in decoherence since the scattered photon is not controlled. Rayleigh scattering, where the final state is the same as the initial state, may also cause decoherence if the scattering amplitudes from the two states that form the superposition are different~\cite{Uys2010}.  

These same decoherence mechanisms arise for ultracold atoms, which are widely used for quantum sensing, computing and simulation. It is instructive to consider the case of a qubit formed by the hyperfine states, $\ket{F,m_F}$, of a ground-state alkali atom confined by a laser whose frequency is below that of the first excited state. Photon scattering leaves the atom in the electronic ground state, so Raman scattering must change $F$ or $m_F$ by re-orienting the spins. Since the light couples only to the spatial degree of freedom, this change must occur through the spin-orbit interaction in the electronic excited states, but that coupling is weak when the detuning of the light ($\Delta$) is large compared to the fine-structure splitting ($\delta_{\rm fs}$). In this limit, the ratio of the rates for Raman scattering to Rayleigh scattering is of order $(\delta_{\rm fs}/\Delta)^2$, which can be very small, especially for light atoms. Rayleigh scattering amplitudes are almost identical for the two hyperfine states when the detuning is large, so decoherence is small for this scattering channel too. Thus, the quantum coherence of an atomic hyperfine qubit is resilient to photon scattering, in some cases surviving for hundreds of photon scattering events~\cite{Cline1994, Ozeri2005, Ozeri2007, Uys2010}. This good fortune does not extend to molecules because photon scattering can change the rotational state of a molecule directly. 

These state-changing processes are related to the vector and tensor parts of the polarizability which are both strongly suppressed in a ground-state alkali atom when the light is far from any resonances. The tensor part is not generally suppressed in a molecule, resulting in a much larger Raman scattering rate. This has consequences for many applications. Cooling to quantum degeneracy requires collisional shielding~\cite{Matsuda2020, Schindewolf2022, Bigagli2024} which relies on the molecules remaining in the same internal state throughout the evaporative cooling stage, which may take many seconds. Applications in quantum computing and simulation rely on the dipole-dipole interactions between molecules which are quite weak, at least when compared with Rydberg atoms, so long coherence times are especially important~\cite{Cornish2024}. Similarly, molecule-based tests of fundamental physics will benefit from the longest possible spin-coherence times~\cite{Fitch2020b}. Knowing the decoherence rate due to photon scattering is important for all these cases. The purpose of this paper is to calculate it.

\section{Formalism}

When light of frequency $\omega_{\rm L}$, intensity $I$, and polarization $\vec{\epsilon}=\sum_p \epsilon_p \hat{e}_p$, is far detuned from resonances, the rate for scattering from initial state $\ket{i}$ to final state $\ket{f}$ is given by the Kramers-Heisenberg formula,
\begin{equation}
    R_{i \rightarrow f} = \frac{I \omega_{\rm s}^3}{6 \pi \epsilon_0^2 \hbar c^4}\sum_{q=-1}^1 \left|\sum_{q'=-1}^1 \alpha_{fi}^{qq'}(\omega_{\rm L})\epsilon_{q'}\right|^2
    \label{eq:R_if}
\end{equation}
where 
\begin{equation}
    \alpha_{fi}^{qq'}(\omega_{\rm L})\!=\!\frac{1}{\hbar}\!\sum_e\! \left[ \frac{\bra{f}d_q\ket{e}\bra{e}d_{q'}\ket{i}}{\omega_{ei}-\omega_{\rm L}-i\gamma_e/2} \!+\! \frac{\bra{f}d_{q'}\ket{e}\bra{e}d_{q}\ket{i}}{\omega_{ef}+\omega_{\rm L}+i\gamma_e/2}\right].
\end{equation}
Here, the sum is over all excited states $\ket{e}$, $\gamma_e$ are the decay rates, $\hbar\omega_{ab}=E_a-E_b$ is the difference in energy between states $a$ and $b$, $\omega_{\rm s}=\omega_{\rm L}-\omega_{fi}$ is the frequency of the scattered photon, and $\vec{d}$ is the electric dipole operator. 

The Raman scattering rate for state $\ket{i}$ is $R^i_{\rm Raman}=\sum_{f \ne i} R_{i \rightarrow f}$, and the Rayleigh scattering rate is $R^i_{\rm Rayleigh}=R_{i \rightarrow i}$. It is convenient to express these rates in terms of the classical Rayleigh scattering rate for an isotropic particle of polarizability $\alpha_0$ and, since we are typically concerned with molecules in optical traps, in terms of the (scalar) trap depth $U=\alpha_0 I/(2 c \epsilon_0)$:
\begin{equation}
    R_{0} = \frac{4\pi^2 \alpha_0^2 I}{3 \epsilon_0^2 \hbar c \lambda^3} = \frac{8\pi^2}{3 \epsilon_0 \hbar \lambda^3}\alpha_0 U.
    \label{eq:R0}
\end{equation}

Suppose we have a qubit ($\ket{0}$, $\ket{1}$) described by a density operator $\rho$. Photon scattering causes the coherence, $\bra{0}\rho\ket{1}$, to decay as $\rho_{01}(t) \propto e^{-\Gamma_{\rm sc}t}$, where $\Gamma_{\rm sc} = \Gamma_{\rm Raman}+\Gamma_{\rm Rayleigh}$ is the total decoherence rate due to both Raman and Rayleigh scattering. These are
\begin{equation}
    \Gamma_{\rm Raman} = \frac{1}{2}\left(R^{0}_{\rm Raman}+R^{1}_{\rm Raman}\right),
    \label{eq:Gamma_Raman}
\end{equation}
and 
\begin{equation}
    \Gamma_{\rm Rayleigh} = \frac{I \omega_{\rm s}^3}{6 \pi \epsilon_0^2 \hbar c^4}\sum_{q=-1}^1 \left|\sum_{q'=-1}^1 \frac{1}{\sqrt{2}}(\alpha_{00}^{qq'} - \alpha_{11}^{qq'})\epsilon_{q'}\right|^2.
    \label{eq:Gamma_Rayleigh}
\end{equation}
For Raman, the decoherence rate is the average Raman scattering rate. For Rayleigh, the decoherence rate is proportional to the square of the difference in Rayleigh scattering amplitudes~\cite{Uys2010}. Throughout this paper, we refer to the decay rate of $\rho_{01}$ as the decoherence rate though, in practice, a large part of the Raman scattering takes the molecule out of the qubit space, appearing as detectable loss.

Equation (\ref{eq:R_if}) can be evaluated directly, but it is more instructive and more convenient to express the result in terms of a polarizability operator in spherical tensor form. To do this, we form the operator $\alpha^{qq'}$ whose matrix elements are $\alpha_{fi}^{qq'}=\bra{f}\alpha^{qq'}\ket{i}$. When $|\omega_{ei}-\omega_{\rm L}|\gg \gamma_e$ for all $e$, which is usually the case, we can neglect the imaginary part in the denominator. Furthermore, since we are interested in cases where $\ket{i}$ and $\ket{f}$ are all within the ground electronic and vibrational manifold, we can use a common energy reference $E_g$ and replace $\omega_{ei}$ and $\omega_{ef}$ by $\omega_{eg}$. With these approximations, we can write
\begin{align}
    \alpha^{qq'} &= d_q \mathcal{R}^{-} d_{q'} + d_{q'} \mathcal{ R}^{+} d_q \nonumber \\
    &= \sum_{K=0}^2 \sum_{P=-K}^K(-1)^P\sqrt{2K+1} 
    \begin{pmatrix}
      1 & 1 & K \\
      q & q' & -P 
    \end{pmatrix}
    z_K\mathcal{A}^{(K)}_P
    \label{eq:alpha_qqprime}
\end{align}
where we have introduced
\begin{equation}
    \mathcal{R}^{\pm}=\frac{1}{\hbar}\sum_e \frac{\ket{e}\bra{e}}{\omega_{eg}\pm\omega_{\rm L}},
\end{equation}
\begin{equation}
    \mathcal{A}^{(K)}_P=(\mathcal{A}^-)^{(K)}_P + (-1)^K(\mathcal{ A}^+)^{(K)}_P,
\end{equation}
and 
\begin{equation}
    (\mathcal{A}^\pm)^{(K)}_P = \frac{1}{z_K}T^{(K)}_P(d,\mathcal{ R}^{\pm}d).
\end{equation}
In this last equation, $T^{(K)}(u,v)$ is the spherical tensor of rank $K$ formed from vectors $\vec{u}$ and $\vec{v}~$\footnote{
$ T^{(K)}_P(u,\!v)\!=\!(-1)^P \sqrt{2K\!+\!1}\!\sum\limits_{p=-1}^1\! \begin{pmatrix}
      1 & 1 & K \\
      p & P\!-\!p & -\!P 
    \end{pmatrix}\!T^{(1)}_p(u) T^{(1)}_{P-p}(v)$}. 
The result is independent of the $z_K$ which are numerical factors introduced for consistency with the normalization used elsewhere in the literature; their values are $z_0=-\sqrt{3}$, $z_1=-\sqrt{2}$, $z_2=\sqrt{3/2}$.

To calculate $R_{i \rightarrow f}$ we use  Eq.~(\ref{eq:R_if}) with $\alpha_{fi}^{qq'}$ determined using Eq.~(\ref{eq:alpha_qqprime}) along with the matrix elements of $\mathcal{A}^{(K)}_P$ between the states of interest. These matrix elements are derived in \cite{Caldwell2020b} and the results for $^{1}\Sigma$ and $^{2}\Sigma$ molecules are reproduced in the Appendix. Conveniently, the matrix elements factorize into a part that depends only on $K,P$ and the quantum numbers of $i$ and $f$, and a frequency-dependent part that contains the sum over excited states but is independent of $i$ and $f$: $\bra{f}\mathcal{A}^{(K)}_P\ket{i} = g(i,f,K,P) \alpha_{K}(\omega_{\rm L})$. The $\alpha_{K}$ are the components of the polarizability in the frame of the molecule. To evaluate them, we introduce the components of the molecule-frame polarizability parallel and perpendicular to the molecular bond,
\begin{align}
    &\alpha_{||}\!=\!\frac{1}{\hbar}\!\sum_j\!\left(\frac{1}{\omega_{j}\!-\!\omega_{\rm L}}\!+\!\frac{1}{\omega_{j}\!+\!\omega_{\rm L}}\right)\left|\bra{X\,\Sigma}d_0\ket{j\,\Sigma} \right|^2|\langle v|v'\rangle|^2, \label{eq:apha_par}\\
    &\alpha_{\perp}\!=\!\frac{1}{\hbar}\!\sum_k\!\left(\frac{1}{\omega_{k}\!-\!\omega_{\rm L}}\!+\!\frac{1}{\omega_{k}\!+\!\omega_{\rm L}}\right)\left|\bra{X\, \Sigma}d_1\ket{k\,\Pi} \right|^2|\langle v|v'\rangle|^2.\label{eq:apha_perp}
\end{align}
Here, the sums run over electronic and vibrational states, $j$ labels the set of excited $\Sigma$ states with frequencies $\omega_j$ above the ground state, $k$ labels the set of excited $\Pi$ states with frequencies $\omega_k$, $v$ and $v'$ are the vibrational quantum numbers of ground and excited states, and we are using components of the dipole operator in the molecule frame. We have explicitly separated the transition dipole moment into its electronic and vibrational parts. Then, for both $^{1}\Sigma$ and $^{2}\Sigma$ molecules\footnote{For a $^{2}\Sigma$ molecule we can distinguish the two spin-orbit manifolds of the $^{2}\Pi_{\Omega}$ states to define $\alpha_{\perp,\Omega}$, with $\alpha_{\perp}=\frac{1}{2}(\alpha_{\perp,\frac{1}{2}}+\alpha_{\perp,\frac{3}{2}})$.}, we have
\begin{align}
    \alpha_0 &= \frac{1}{3}(\alpha_{||}+2\alpha_{\perp}),\\
    \alpha_2 &= \frac{2}{3}(\alpha_{||}-\alpha_{\perp}).
\end{align}
For a $^{1}\Sigma$ molecule, neglecting singlet-triplet mixing, we have $\alpha_1 =0$. For a $^{2}\Sigma$ molecule, we account for the fine-structure splitting between excited $^{2}\Pi_{1/2}$ and $^{2}\Pi_{3/2}$ states, obtaining
\begin{equation}
    \alpha_1  = \frac{1}{2}(\beta_{\perp,\frac{1}{2}} - \beta_{\perp,\frac{3}{2}}), 
    \label{eq:alpha1}
\end{equation}
where
\begin{align}
    \beta_{\perp,\Omega} \! = \! \frac{1}{\hbar}\!\sum_{k}&\left(\frac{1}{\omega_{k,\Omega}\!+\!\omega_{\rm L}}\!-\!\frac{1}{\omega_{k,\Omega}\!-\!\omega_{\rm L}}\right)\times \nonumber\\
    &\left|\bra{X\,^{2}\Sigma}d_1\ket{k\,^{2}\Pi_{\Omega}} \right|^2|\langle v|v'\rangle|^2. \label{eq:beta}
\end{align}

The scalar part of $\mathcal{A}$ contributes to Rayleigh scattering but not to Raman scattering since $\bra{f}\mathcal{A}^{(0)}_0\ket{i} = 0$ for $f\ne i$. The vector and tensor parts are responsible for Raman scattering. When the detuning of the light is large compared to any spin-orbit splitting, $\alpha_1$ is small and Raman scattering will be dominated by the rank-2 part of $\mathcal{A}$. 

Because our ground manifold containing $\ket{i}$ and $\ket{f}$ is formed from rotational and hyperfine levels of a single vibrational state, we are neglecting Raman scattering between different vibrational states. As we will see, this is usually a very good approximation, but it breaks down at frequencies where the Raman scattering is very small, which is a regime we are interested in. Fortunately, because the rotational structure is independent of vibrational state, the formalism is easily extended to scattering into other vibrational states. We define the quantities $\alpha_{||,\perp,0,1,2}^{vv''}$ to represent \textit{transition polarizabilities} for scattering from $X(v)$ to $X(v'')$, with an exact analogy to the state polarizabilites defined by Eqns.~(\ref{eq:apha_par})-(\ref{eq:alpha1}). Specifically,
\begin{align}
    \alpha_{||}^{vv''}\!=\!\frac{1}{\hbar}\!\sum_j&\!\left(\frac{1}{\omega_{j}\!-\!\omega_{\rm L}}\!+\!\frac{1}{\omega_{j}\!+\!\omega_{\rm L}}\right)\times\nonumber \\
    &\left|\bra{X\,\Sigma}d_0\ket{j\,\Sigma} \right|^2\langle v|v'\rangle\langle v'|v''\rangle,
    \label{eq:alpha_par_vib}
\end{align}
where, as before, $v'$ is the vibrational component of the intermediate state $j$. An identical modification to Eqs.~(\ref{eq:apha_perp}) and (\ref{eq:beta}) is applied to form $\alpha_{\perp}^{vv''}$ and $\beta_{\perp}^{vv''}$, from which we construct $\alpha_{0,1,2}^{vv''}$. All subsequent steps to calculate the vibrational Raman scattering are identical to the ones described above\footnote{Equating the energies of initial and final states is not such a good approximation here, but still typically good enough, while $\omega_s = \omega_{\rm L}-\omega_{vv''}$ is straightforward to include.}. Importantly, the overlap integrals in Eq.~(\ref{eq:alpha_par_vib}) are signed quantities, so they are not obtainable from Franck-Condon factors. Since $v'' \ne v$, at least one of the two overlap integrals forming the product is non-diagonal. Moreover, when the detuning is large compared to the vibrational spacing, there are near cancellations between terms of different $v'$ in the sum. This is easily seen by taking the limit where the $v'$ dependence of $\omega_j$ is neglected. Then, we can use the closure relation $\sum_{v'}\ket{v'}\bra{v'}=1$ and the orthogonality $\langle v|v'' \rangle=0$ to see that $\alpha_{||}^{vv''}$ goes to zero in this limit. For these reasons, vibrational Raman scattering is typically small.

\section{Comparison with the light shift}\label{sec:light_shift_comparison}

It is interesting to draw parallels between the photon scattering rate and the light shift, since they both depend on the same operator, $\mathcal{A}$. Experiments with ultracold molecules in optical traps usually seek conditions where the light shift is identical for all states of interest, since differential shifts lead to rapid dephasing of superposition states. This amounts to eliminating the rank-1 and rank-2 parts since they are the state-dependent parts. This can sometimes be done by finding a \textit{magic wavelength} where both $\alpha_1$ and $\alpha_2$ are zero, or very nearly so. Except for the heaviest molecules, $\alpha_1$ is very small away from resonances, so it is typically $\alpha_2$ that needs to be zeroed. An alternative is to do an experiment exclusively in the rotational ground state, $N=0$, which has no tensor light shift since the diagonal matrix elements of $\mathcal{A}^{(2)}$ are zero\footnote{Hyperfine interactions can lead to non-zero tensor light shifts in the rotational ground state, but this is usually very small and we will neglect it for the present discussion} even when $\alpha_2$ is not. When this restriction is unwelcome, one might instead be able to find a \textit{magic polarization} where the light shifts of the states of interest are almost equal. The rank-$K$ part of the light shift is proportional to a scalar product of $\mathcal{A}^{(K)}$ with $[\vec{\epsilon} \otimes \vec{\epsilon}\,^{*}]^{(K)}$. For $K=1$, this is the cross product $-i(\vec{\epsilon} \times \vec{\epsilon}\,^*)$ which is zero for any linear polarization. The rank-2 case is more complicated, except for the special case where rotational states $\ket{N,m_N}$ are good eigenstates and the energy difference between states of the same $N$ but different $m_N$ is large compared to the light shift. In that case, matrix elements off-diagonal in $m_N$ have little influence, and the tensor light shift is determined mainly by $\mathcal{A}^{(2)}_0 [\vec{\epsilon} \otimes \vec{\epsilon}\,^*]^{(2)}_0$. For light linearly polarized at angle $\theta$ to the $z$-axis, this component of the polarization tensor is the Legendre polynomial $P_{2}(\cos\theta)$ which is zero when $\cos^2\theta = 1/3$. Away from this special case, a specific choice of polarization, laser intensity and applied magnetic field can sometimes be found where the light shift is equal for the states of interest. 

Can any of these strategies be carried over to suppress decoherence due to photon scattering? For example, it is natural to wonder whether Raman scattering is suppressed in the ground rotational state. It is not. While the diagonal elements of $\mathcal{A}^{(2)}$ are zero, Raman scattering is determined by the off-diagonal elements which are not zero. Similarly, we might wonder whether the contribution of $\alpha_1$ to Raman scattering is suppressed when the light is linearly polarized. It is not. The vector part of the Raman scattering into a mode of polarization $\vec{\eta}$ is proportional to $\vec{\epsilon} \times \vec{\eta}\,^*$, which is not zero since we have to sum over all possible polarizations of the scattered photon. Linear polarization does not eliminate the rank-1 contribution to Rayleigh scattering either, unless the polarization is parallel to the quantization axis. Similarly, there is no magic angle that will eliminate the tensor contribution to Raman scattering. By contrast, finding a magic wavelength where both $\alpha_1$ and $\alpha_2$ are near zero \textit{does} suppress decoherence due to photon scattering, as well as suppressing the dephasing by the light shift, so this is the best strategy to ensure long coherence times.

\section{Applications}

\subsection{$^{1}\Sigma$ molecules}

Consider a $^{1}\Sigma$ molecule in rotational state $\ket{N,m_N}$ and assume that the light is either pure linear or pure circular with respect to the $z$-axis so that only one value of $q'$ contributes to the summation. Using Eqs.~(\ref{eq:R_if}), (\ref{eq:R0}), (\ref{eq:alpha_qqprime}) and (\ref{eq:me_singlet}), the Rayleigh scattering rate is
\begin{equation}
    R^{(N,m_N)}_{\rm Rayleigh}=R_{0}\left(1 +\frac{\alpha_2}{\alpha_0} f_{q'}^{(N,m_N)}\right)^2,
    \label{eq:singlet_Rayleigh}
\end{equation}
where
\begin{equation}
    f_{q'}^{(N,m_N)}=\left(1-3q'^2/2 \right)\langle C^{(2)}_0\rangle
\end{equation}
and
\begin{equation}
\langle C^{(2)}_0\rangle \!=\! \bra{N,m_N} C^{(2)}_0\ket{N,m_N}\!=\!\frac{N(N\!+\!1)\!-\!3m_N^2}{(2N\!-\!1)(2N\!+\!3)}
\end{equation}
is the expectation value of the (renormalized) rank-2 spherical harmonic. The Raman scattering rate is
\begin{equation}
    R^{(N,m_N)}_{\rm Raman}= R_{0} \frac{\alpha_2^2}{\alpha_0^2}g_{q'}^{(N,m_N)}
    \label{eq:singlet_Raman}
\end{equation}
where
\begin{equation}
    g_{q'}=\frac{1}{2}(1+2f_{q'})(1-f_{q'}).
\end{equation}
The explicit expressions for $g_{q'}$ are
\begin{align}
    g_0^{(N,m_N)}&=\frac{9}{16}\left(\!1\!-\!\left(\frac{4m_N^2\!-\!1}{(2N\!-\!1)(2N\!+\!3)}\right)^2 \right),\\
    g_{\pm1}^{(N,m_N)}&=\frac{9}{16}\left(\!1\!-\!\left(\!\frac{2N(N\!+\!1)\!-\!(2m_N^2\!+\!1)}{(2N\!-\!1)(2N\!+\!3)}\!\right)^2 \right).
\end{align}
Note that $g_{q'}^{(0,0)}=1/2$ for all polarizations. It is also interesting to note that $g_0^{(N,N)} \rightarrow 9/(8N)$ at large $N$, so Raman scattering rates become small for large-$N$ stretched states in $\pi$-polarized light.

By analogy to the above, we obtain the vibrational Raman scattering from $v=0$ to $v''$ as
\begin{equation}
    R_{\rm Raman}^{0v''}=\frac{R_0}{\alpha_0^2}\left[\left(\alpha_0^{0v''} +\alpha_2^{0v''} f_{q'}\right)^2 + (\alpha_2^{0v''})^2 g_{q'} \right].
    \label{eq:R_Raman_vib}
\end{equation}
This part is usually negligible.

Now suppose we form a coherent superposition of neighbouring rotational states, specifically $\ket{N,0}$ and $\ket{N+1,0}$. This is an important case since many applications of ultracold molecules rely on coherences of this type. The decoherence rates are
\begin{equation}
    \Gamma_{\rm Raman} = \frac{1}{2}\left(R^{(N,0)}_{\rm Raman}+R^{(N+1,0)}_{\rm Raman}\right),
    \label{eq:Gamma_Raman_singlet}
\end{equation}
and
\begin{equation}
    \Gamma_{\rm Rayleigh} = R_{0}\frac{\alpha_2^2}{\alpha_0^2}\frac{(h_{q'}^N)^2}{2},
    \label{eq:Gamma_Rayleigh_singlet}
\end{equation}
where
\begin{align}
    &h_{q'}^N=f_{q'}^{(N,0)}-f_{q'}^{(N+1,0)}.
\end{align}
The explicit expression for $h_{q'}$ is
\begin{equation}
    h_{0}^N\!=\!\left(\frac{6(N+1)}{(2N\!-\!1)(2N\!+\!1)(2N\!+\!3)(2N\!+\!5)} \right).
\end{equation}
Finally, we arrive at the total decoherence rate,
\begin{equation}
    \Gamma_{\rm sc} = R_0 \frac{\alpha_2^2}{\alpha_0^2}\left[\frac{1}{2}\!+\!\frac{1}{4}\left(f_{q'}^{(N,0)}\!+\!f_{q'}^{(N+1,0)}\right)\!-\!f_{q'}^{(N,0)}f_{q'}^{(N+1,0)}\right],
    \label{eq:Gamma_sc_singlet}
\end{equation}
where we have neglected the small vibrational part.

Even though $\alpha_0$ and $\alpha_2$ both contribute to the Rayleigh scattering rate, the decoherence rate due to Rayleigh scattering is proportional only to $\alpha_2^2$, just like the decoherence due to Raman scattering. For linear polarization ($q'=0$) the total decoherence rate is $\frac{3}{5}R_{0}\alpha_2^2/\alpha_0^2$ for a superposition of $N=0$ and $N=1$, and rapidly approaches $\frac{9}{16}R_{0}\alpha_2^2/\alpha_0^2$ as $N$ increases. 

As noted in Sec.~\ref{sec:light_shift_comparison}, decoherence due to photon scattering will be strongly suppressed at magic wavelengths where $\alpha_2=0$, or equivalently where $\alpha_{||}=\alpha_{\perp}$. This is exactly the same condition that eliminates the rotational-state-dependence of the light shift, which is necessary to achieve long coherence times in optical traps and has been studied for the $^{1}\Sigma$ molecules NaK, RbCs and NaRb~\cite{Bause2020, He2021, Lin2021, Guan2021, Kotochigova2024, Gregory2024}. For these molecules, there is a magic wavelength close to a weakly-allowed transition from $X ^{1}\Sigma$ to $b ^{3}\Pi_0$ where $\alpha_2 = 0$. Thus, the problem of decoherence due to photon scattering is already mitigated for these molecules. Note that there can still be vibrational Raman scattering here which, though small, is likely to be the dominant contribution at the magic wavelength especially when the light is close to a resonance. It is also worth noting that the transition to $b ^{3}\Pi_0$ relies on singlet-triplet mixing and that this may introduce a significant vector polarizability near these resonances. This would be interesting to explore. 

\subsection{$^{2}\Sigma$ molecules}

With non-zero electronic spin, the vector polarizability also contributes to the scattering rate, introducing spin-decoherence as well as rotational decoherence. Consider a $^{2}\Sigma$ molecule with nuclear spin $I=1/2$, and label states as $\ket{N,J,F,m_F}$. We calculate the scattering rates using Eqs.~(\ref{eq:R_if}), (\ref{eq:alpha_qqprime}) and (\ref{eq:me_doublet}). In general, $\alpha_{0,1,2}$ may all contribute to the Rayleigh scattering rate,
\begin{equation}
     R_{\rm Rayleigh}=R_{0}\left(1+c_1\frac{\alpha_1}{\alpha_0} +c_2\frac{\alpha_2}{\alpha_0} \!\right)^2,
\end{equation}
where the coefficients $c_{1,2}$ depend on the quantum numbers and on $q'$. We have $c_1=0$ when $q'=0$ or $m_F=0$, and $c_2=0$ when $J=1/2$.

Both rotational qubits and hyperfine qubits have been used in experiments with $^{2}\Sigma$ molecules~\cite{Bao2023, Holland2023, Holland2025}. The natural hyperfine qubit is formed from $\ket{0}=\ket{0,1/2,0,0}$ and $\ket{1}=\ket{0,1/2,1,0}$, and the rotational qubit from $\ket{1}$ and $\ket{2}=\ket{1,1/2,0,0}$. For all three of these qubit states, 
\begin{align}
    &R_{\rm Rayleigh} = R_0,\label{eq:R_Rayleigh_simple}\\
    &R_{\rm Raman}=\frac{R_{0}}{\alpha_0^2} \left(\frac{2}{3}\alpha_1^2 + \frac{1}{2}\alpha_2^2 \right),\label{eq:R_Raman_simple}\\
    &R_{\rm Raman}^{0v''}=\frac{R_{0}}{\alpha_0^2} \left((\alpha_0^{0v''})^2+\frac{2}{3}(\alpha_1^{0v''})^2 + \frac{1}{2}(\alpha_2^{0v''})^2 \right),\label{eq:R_Raman_vib_simple}
\end{align}
irrespective of the polarization of the light. For the initial state $\ket{0}$, the tensor part couples the molecule to the $F=2$ components of $N=2$, while the vector part couples to the $F=1$ component of both $N=0$ and $N=2$. For the initial state $\ket{2}$, the tensor part couples to the $F=2$ component of both $N=1$ and $N=3$, while the vector part couples only to the $F=1$ components of $N=1$. Since the Rayleigh amplitudes and Raman rates are equal for all 3 qubit states, the decoherence rate is equal to the total Raman scattering rate, with no contribution from Rayleigh scattering. Again, to suppress this decoherence, $\alpha_{1,2}$ and $\alpha_{0,1,2}^{0v''}$ all need to be small.

\begin{figure*}
    \centering
    \includegraphics[width=1\linewidth]{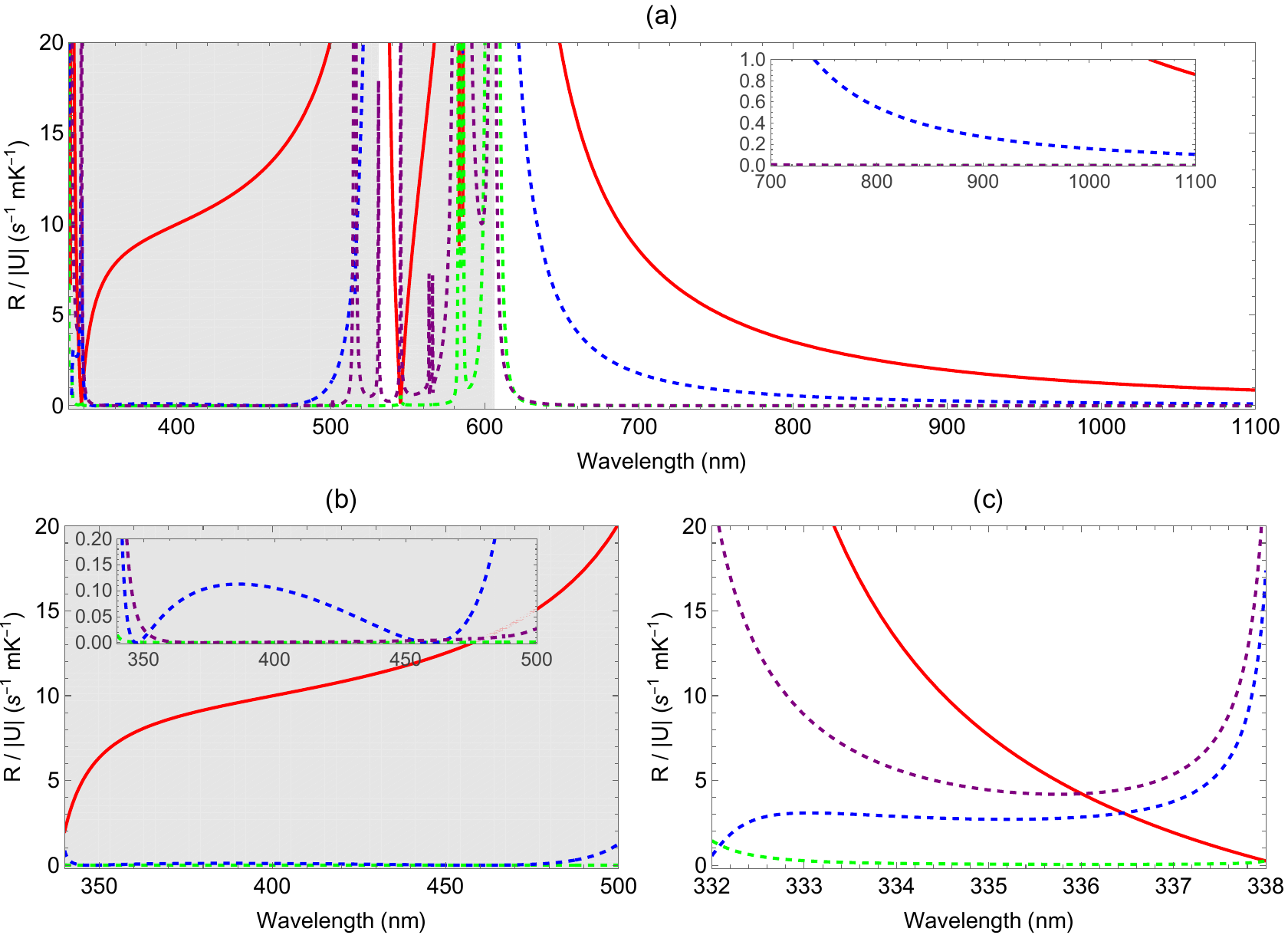}
    \caption{Scattering rates for a ground state CaF molecule in a trap of depth 1~mK. Solid red lines: Rayleigh scattering. Dashed lines: Raman scattering, broken down into vector (green), tensor (blue) and vibrational (purple) parts. Unshaded areas have $\alpha_0>0$ and shaded areas have $\alpha_0 <0$. (a) 330-1100~nm. (b) 340-500~nm. (c) 332-338~nm. Insets magnify the low $R$ regions.}
    \label{fig:CaF}
\end{figure*}

In contrast to $^{1}\Sigma$ molecules formed from alkali atoms, there have not been many studies of magic wavelength conditions for $^{2}\Sigma$ molecules. Here, we consider the case of CaF as a typical laser-coolable $^{2}\Sigma$ molecule. We determine $\alpha_{0,1,2}$ using Eqs.~(\ref{eq:apha_par})-(\ref{eq:beta}) and $\alpha_{0,1,2}^{0v''}$ from the equivalent expressions (Eq.~(\ref{eq:alpha_par_vib}) etc.). We include $A{}^2\Pi$, $B{}^2\Sigma$, $C{}^2\Pi$ and $D{}^2\Sigma$ in the sums over excited states, using spectroscopic constants from \cite{Kaledin1999,Gittins1993,Prasad1969} and electronic transition dipole moments from \cite{Wall2008, Dagdigian1974, Raouafi2001}. The vibrational overlap integrals are calculated by constructing Rydberg-Klein-Rees potential energy curves and solving the Schr\"odinger equation to find the vibrational eigenfunctions for these potentials. 

Figure \ref{fig:CaF} shows the Rayleigh rate $R_{\rm Rayleigh}/|U|$, rotational Raman rate $R_{\rm Raman}/|U|$, and vibrational Raman rate $R_{\rm Raman}^{01}/|U|$, all per unit trap depth, for ground-state CaF over various wavelength ranges. In these plots, the unshaded regions have $\alpha_0 > 0$ meaning that the molecule is attracted to high intensity light, while the shaded regions are the opposite. Contributions to the vibrational Raman rate from higher $v''$ are found to be negligible over the whole wavelength range explored here.

Figure \ref{fig:CaF}(a) shows the scattering rates across a wide range of wavelengths that includes the transitions from $X{}^2\Sigma$ to $A{}^2\Pi$, $B{}^2\Sigma$, $C{}^2\Pi$ and $D{}^2\Sigma$. The long wavelength range beyond 700~nm is frequently used for optical trapping. At $\lambda=1064$~nm, a common wavelength for optical dipole traps, the total Rayleigh and Raman rates are 0.98 and 0.12~s$^{-1}$~mK$^{-1}$. At a trap depth of 100~$\mu$K, which is sufficient for most applications, the coherence time is 83~s and the motional heating rate\footnote{The heating rate used here is $2R E_{\rm r}/k_{\rm B}$ where $R$ is the total scattering rate and $E_{\rm r}=(\hbar k)^2/(2m)$ is the recoil energy.} due to the recoil of scattered photons is 31~nK/s. In this example, a molecule scatters an average of 9.2 photons before its coherence is lost. Experiments with single CaF molecules in tweezer traps~\cite{Anderegg2019, Bao2023, Holland2023} have used $\lambda \approx 780$~nm. Here, the Rayleigh and Raman rates are 4.1 and 0.66~s$^{-1}$~mK$^{-1}$, so at 100~$\mu$K trap depth the coherence time is 15~s and the motional heating rate is 252~nK/s.

Figure \ref{fig:CaF}(b) shows the scattering rates between 340 and 500~nm. The total Raman rate is below 2.3~s$^{-1}$~mK$^{-1}$ and the Rayleigh rate below 20~s$^{-1}$~mK$^{-1}$ throughout this region. The Raman rate is below 0.13~s$^{-1}$~mK$^{-1}$ between 345 and 480~nm, with two minima arising from zero crossings of $\alpha_2$ near 347~nm and 460~nm. $\alpha_1$ is also tiny at these wavelengths, so the residual Raman scattering at the minima is dominated by the vibrational scattering to $X ^{2}\Sigma (v''=1)$. With this contribution included, the two minima are at $\lambda=350.8$~nm, where the total Raman rate is 0.035~s$^{-1}$~mK$^{-1}$ and the Rayleigh rate is 6.5~s$^{-1}$~mK$^{-1}$, and at $\lambda=459.1$~nm, where the Raman rate is 0.005~s$^{-1}$~mK$^{-1}$ and the Rayleigh rate is 13.2~s$^{-1}$~mK$^{-1}$. Thus, this region offers extremely long coherence times against photon scattering, negligible vector and tensor light shifts, and low motional heating rates. However, $\alpha_0 <0$ in this region so confinement requires an optical lattice or a hollow beam or bottle beam trap~\cite{Kuga1997, Barredo2020}. In such a trap, the molecules are confined around the intensity minimum, so the scattering rates will be even lower than the ones estimated here.

Figure \ref{fig:CaF}(c) shows the scattering rates between 332 and 338~nm. The scattering rates remain relatively low in this region, despite the $1/\lambda^3$ scaling, and the vibrational contribution starts to dominate over the rotational contribution to the Raman scattering. For example, at $\lambda = 336$~nm, the Rayleigh rate is 4.2~s$^{-1}$~mK$^{-1}$ and the total Raman rate is 7.1~s$^{-1}$~mK$^{-1}$, with vibrational Raman scattering making up about 60\% of this. Optical trapping using standard Gaussian beams is possible here, since $\alpha_0>0$, and at the diffraction limit the confinement would be exceptionally tight due to the small wavelength. This could be very beneficial for applications in quantum simulation and computing that rely on dipole-dipole interactions between neighbouring molecules, with the potential to increase the interaction strength by an order of magnitude compared to tweezer traps in the infra-red. The benefit of the increased interaction strength may outweigh the shorter coherence time which would still reach about 1.4~s in a trap of depth 100~$\mu$K.

\section{Conclusions}

This paper provides the recipe to calculate Rayleigh and Raman scattering rates and the decoherence they induce, for $^{1}\Sigma$ and $^{2}\Sigma$ diatomic molecules. First, calculate $\alpha_0$, $\alpha_1$ and $\alpha_2$ using Eqs.~(\ref{eq:apha_par})-(\ref{eq:beta}). At laser frequencies below the first electronic excited state, it is often sufficient to include only the lowest-lying $\Sigma$ and $\Pi$ states in the sums over excited states. At higher laser frequencies, data for more excited states may be needed. If $\alpha_{1,2}$ are particularly small, it may also be necessary to calculate $\alpha_{0,1,2}^{vv''}$ using the same methods. Next, for your choice of initial state $\ket{i}$, evaluate the matrix elements $\bra{f}\alpha^{qq'}\ket{i}$ using Eq.~(\ref{eq:alpha_qqprime}) and either Eq.~(\ref{eq:me_singlet}) or Eq.~(\ref{eq:me_doublet}). The scattering rates can then be determined using Eq.~(\ref{eq:R_if}). For the Raman rate, the sum over final states $\ket{f}$ only needs to include rotational states that differ from $\ket{i}$ by 0 or 2 units. For any choice of qubit within the molecule, the decoherence rates due to Raman and Rayleigh scattering are found from Eqs.~(\ref{eq:Gamma_Raman}) and (\ref{eq:Gamma_Rayleigh}). For a $^{1}\Sigma$ molecule in rotational state $\ket{N,m_N}$, explicit formulae for the Rayleigh and Raman rates are given by Eqs.~(\ref{eq:singlet_Rayleigh}), (\ref{eq:singlet_Raman}) and (\ref{eq:R_Raman_vib}). For a qubit formed from neighbouring rotational states, the total decoherence rate is given by Eq.~(\ref{eq:Gamma_sc_singlet}). For $^{2}\Sigma$ molecules, explicit expressions for the rates are given only for the most natural qubit states, Eqs.~(\ref{eq:R_Rayleigh_simple}), (\ref{eq:R_Raman_simple}) and (\ref{eq:R_Raman_vib_simple}).

Both Raman and Rayleigh scattering contribute to decoherence, but in both cases the decoherence comes only from $\alpha_1$ and $\alpha_2$. While there are many strategies to minimize dephasing due to differential light shifts in an optical trap, the only effective way to suppress decoherence by photon scattering is to make $\alpha_1$ and $\alpha_2$ as small as possible. Then, the decoherence rate can be far smaller than the total scattering rate, with a floor set by scattering to other vibrational states whose rate is determined by $\alpha_{0,1,2}^{vv''}$. The conditions needed for this are known for several bialkali molecules. For CaF, we estimate an extremely low decoherence rate between 345 and 480~nm. However, $\alpha_0 <0$ in this wavelength range, so the molecules would need to be confined in a lattice or a hollow beam or bottle beam trap. 

It is common to make optical traps at laser frequencies well below the first electronic excited state. Here, $\alpha_1$ is usually negligible, whereas $\alpha_2$ is often similar in size to $\alpha_0$. In this case, the decoherence rate due to photon scattering is much larger than for a ground-state alkali atom, and that is potentially a problem for applications of ultracold molecules. Fortunately, the rates are not too high in practice. Taking ground-state CaF in a trap of depth 100~$\mu$K as an example, the coherence time from photon scattering alone is 83~s at $\lambda=1064$~nm, falling to 15~s at $\lambda=780$~nm. Such coherence times are likely to be long enough for studying the behaviour of quantum degenerate molecular gases, simulating many-body quantum physics with arrays of molecules, computing with molecules, and using molecules to measure symmetry-violating physics beyond the Standard Model.

\acknowledgements
I am grateful to Mehedi Hasan and Chi Zhang for discussions on decoherence in atomic qubits that helped inspire this work. After completing this paper, I used Claude Opus 5.5 to verify the accuracy and consistency of the work and implemented some of the model's suggestions that I judged would improve clarity and completeness. The work has been supported by EPSRC through grants EP/W00299X/1, EP/V011499/1, EP/Z535898/1 and UKRI2226.

\onecolumngrid

\appendix*

\section{Matrix elements of the polarizability operator}

For a $^{1}\Sigma$ molecule with rotational states $\ket{N,m_N}$, the matrix elements of $\mathcal{A}^{(K)}_P$ are
\begin{align}
    \bra{N',m_N'}\mathcal{A}^{(K)}_P\ket{N,m_N} &\!=\! (-1)^{m_N'}\!\sqrt{(2N\!+\!1)(2N'\!+\!1)} 
    \begingroup 
    \setlength\arraycolsep{1pt}
    \begin{pmatrix}
      N' & K & N \\
      -m_N' & P & m_N 
    \end{pmatrix}
     \begin{pmatrix}
      N' & K & N \\
      0 & 0 & 0 
    \end{pmatrix} \alpha_K.
    \endgroup
    \label{eq:me_singlet}
\end{align}

For a $^{2}\Sigma$ molecule, states are labelled $\ket{N,S,J,I,F,m_F}$, where the quantum numbers are the rotational angular momentum, electron spin, total electronic angular momentum, nuclear spin, total angular momentum and its projection onto the $z$-axis. The matrix elements are
\begin{align}
    &\bra{N',S,J',I,F',m_F'}\mathcal{A}^{(K)}_P\ket{N,S,J,I,F,m_F} \
    = ((-1)^{N+N'}+1)(-1)^{F'-m_F'+F-J'+J+I+1/2} \nonumber \\
    & \times \sqrt{(2F+1)(2F'+1)(2N+1)(2N'+1)(2J+1)(2J'+1)} \nonumber\\
    & \times
    \begin{Bmatrix}
      J' & F' & I \\
      F & J & K 
    \end{Bmatrix}
    \begin{pmatrix}
      F' & K & F \\
      -m_F' & P & m_F 
    \end{pmatrix}
    \begin{pmatrix}
      J & 1/2 & N \\
      -1/2 & 1/2 & 0 
    \end{pmatrix} \begin{pmatrix}
      J' & 1/2 & N' \\
      -1/2 & 1/2 & 0 
    \end{pmatrix}
    \begin{pmatrix}
      J' & K & J \\
      -1/2 & 0 & 1/2 
    \end{pmatrix} \alpha_K.
    \label{eq:me_doublet}
\end{align}
\twocolumngrid

\bibliography{references}

\end{document}